\documentclass[10pt,conference]{IEEEtran}
\IEEEoverridecommandlockouts
\usepackage{cite}
\usepackage{amsmath,amssymb,amsfonts}
\usepackage{algorithmic}
\usepackage{graphicx}
\usepackage{textcomp}
\usepackage{booktabs}
\usepackage{tikz}
\usepackage[linesnumbered,ruled,vlined]{algorithm2e}
\usepackage{pgfplots}
\usepackage{caption}
\usepackage{siunitx}
\usepackage{multirow}
\usepackage{subcaption}
\usepackage[left=1.62cm,right=1.62cm,top=1.9cm]{geometry}
\def\BibTeX{{\rm B\kern-.05em{\sc i\kern-.025em b}\kern-.08em
    T\kern-.1667em\lower.7ex\hbox{E}\kern-.125emX}}
    
 \def\BibTeX{{\rm B\kern-.05em{\sc i\kern-.025em b}\kern-.08em T\kern-.1667em\lower.7ex\hbox{E}\kern-.125emX}}
\begin{document}

\title{Fast Butterfly-Core Community Search For Large Labeled Graphs}

\author{\IEEEauthorblockN{JiaYi Du\textsuperscript }
	\IEEEauthorblockA{School of Computer and Information Engineering \\
		Central South University of Forestry and Technology\\
		Changsha, China \\
		dujiayi@csuft.edu.cn}
	\and
	\IEEEauthorblockN{YingHao Wu}
	\IEEEauthorblockA{School of Computer and Information Engineering \\
		Central South University of Forestry and Technology\\
		Changsha, China \\
		wuyinghao@csuft.edu.cn}
	\and
	\IEEEauthorblockN{\hspace{3em}Wei Ai}
	\IEEEauthorblockA{\hspace{4em}School of Computer and Information Engineering \\
		\hspace{4em}Central South University of Forestry and Technology\\
		\hspace{4em}Changsha, China \\
		\hspace{4em}aiwei@hnan.edu.cn}
	\and
	\IEEEauthorblockN{\hspace{-1em}Tao Meng\textsuperscript{*}}
	\IEEEauthorblockA{School of Computer and Information Engineering\\
		Central South University of Forestry and Technology\\
		Changsha, China \\
		mengtao@hnan.edu.cn}
	\and
	\IEEEauthorblockN{\hspace{4em}CanHao Xie}
	\IEEEauthorblockA{\hspace{4em}School of Computer and Information Engineering \\
		\hspace{4em}Central South University of Forestry and Technology\\
		\hspace{4em}Changsha, China \\
		\hspace{4em}Xiecanhao@csuft.edu.cn,}
	\and
	\IEEEauthorblockN{\hspace{2em}Keqin Li}
	\IEEEauthorblockA{\hspace{3em}Department of Computer Science \\
		\hspace{3em}State University of New York\\
		\hspace{3em}New Paltz, New York 12561, USA \\
		\hspace{4em}lik@newpaltz.edu}
	\thanks{*Tao Meng is the corresponding author.}
	
}
\maketitle

\begin{abstract}
Community Search (CS) aims to identify densely interconnected subgraphs corresponding to query vertices within a graph. However, existing heterogeneous graph-based community search methods need help identifying cross-group communities and suffer from efficiency issues, making them unsuitable for large graphs. This paper presents a fast community search model based on the Butterfly-Core Community (BCC) structure for heterogeneous graphs. The Random Walk with Restart (RWR) algorithm and butterfly degree comprehensively evaluate the importance of vertices within communities, allowing leader vertices to be rapidly updated to maintain cross-group cohesion. Moreover, we devised a more efficient method for updating vertex distances, which minimizes vertex visits and enhances operational efficiency. Extensive experiments on several real-world temporal graphs demonstrate the effectiveness and efficiency of this solution.
\end{abstract}

\begin{IEEEkeywords}
Community Search, Butterfly-Core, Labeled Graph, Heterogeneous Graphs, Random Walk with Restart
\end{IEEEkeywords}

\section{Introduction}\label{A}
Graphs are essential for depicting social, academic, expertise, and IT professional networks \cite{33, 34, 35, 36, 37, 38, 39, 40, 41, 42, 43, 44}. These networks are labeled graphs, where vertices have labels, such as roles in IT professional networks. A vital feature of these graphs is the emergence of cross-group communities, as illustrated in Figure 1. These arise when closely collaborating groups with different labels form, as seen in cross-role business collaborations within IT professional networks. In this context, we focus more on heterogeneous graphs. Heterogeneous graphs have vertices with different labels, each vertex having a single label type. In contrast, homogeneous graphs either lack vertex labels or have all vertices with the same label.

\begin{figure}[htb]
	\centering
	\includegraphics[width=0.46\textwidth, height=0.15\textheight]
	{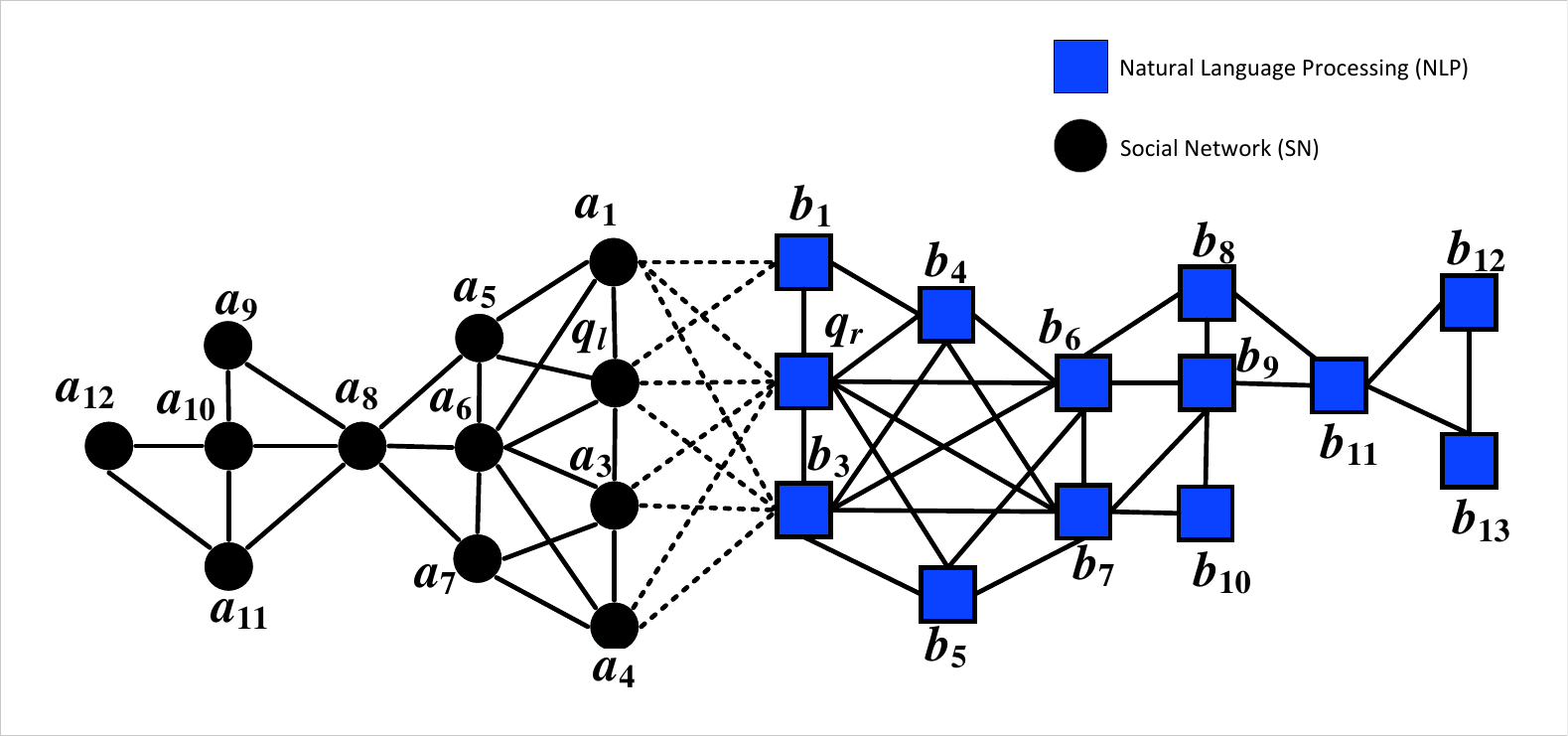}
	\caption{An instance of a labeled graph denoted as \textit{G} within IT professional networks showcasing two labels, namely NLP and SN. The solid edges (and dashed edges) represent interactions between homogeneous vertices (vertices of the same type) and heterogeneous vertices (vertices of different types).}
	\label{fig:figure1}
\end{figure}
In recent years, research on community search (CS) tasks has witnessed the emergence of various models for identifying densely connected subgraphs, such as the \textit{k}-core [2], \textit{k}-truss [3], quasi-clique [4], and densest subgraph [5], within the context of CS tasks. While these community search models primarily focus on identifying homogeneous communities [6], they often overlook the semantic information carried by vertices and edges. The butterfly-core community search [1] aims to locate densely connected cross-group communities, yet this approach also presents certain limitations. For instance, the greedy leader vertex identifying method's emphasis on high butterfly-degree leader vertices within a 3-hop range focuses too much on interactions among heterogeneous community vertices, neglecting interactions among homogeneous ones. This method overlooks vertices within the community with significant influence and suitable butterfly degrees, making the leader vertex search time-consuming. Furthermore, the BCC search model necessitates the diameter reduction of subgraphs to obtain a small, dense community. However, during graph reduction, checking the distance from each vertex to the query vertex resulted in unnecessary time consumption.

Motivated by the above, we present a fast community search solution founded on a butterfly-core community structure. Specifically, since the \textit{Random Walk with Restart}(RWR) algorithm captures intimacy scores between vertices, we propose a vertex comprehensive scoring method—melding RWR's intimacy score with the vertex's butterfly degree to assess vertex significance within the community. In graph reduction, we devised a novel leader update method using vertex comprehensive scores. Motivating by BFS, we introduced a new technique to enhance BCC retrieval efficiency by updating vertex distances.

To summarize, our contributions are as follows:

$\bullet$ We introduce a comprehensive vertex scoring approach that combines the RWR algorithm's vertex score with the butterfly degree to swiftly identify and update the leader vertex's significance within a community.

$\bullet$ We develop a more efficient approach to update the leader vertex based on the vertex's comprehensive score.

$\bullet$ We design a fast method for updating vertex distances based on the concept of the BFS algorithm.
\begin{figure}[htb]
	\centering
	\includegraphics[width=0.35\textwidth, height=0.1\textheight]{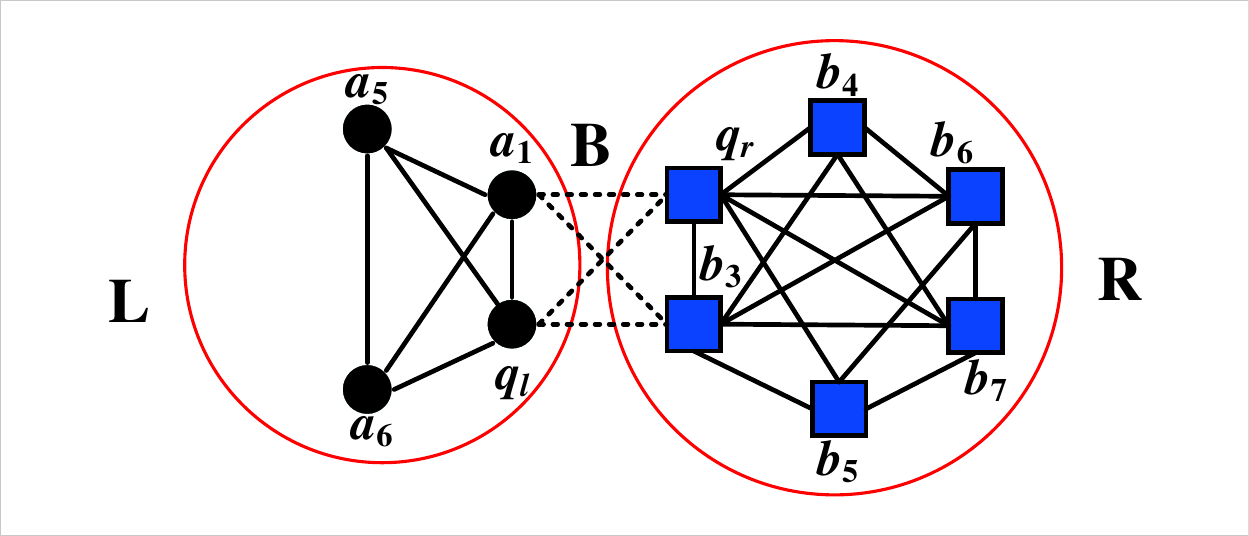}
	\caption{An instance of a butterfly-core community \textit{$G_H$}. The bowtie \textit{B} emerges from the collection of dashed edges connecting two labeled groups.}
	\label{fig:figure2}
\end{figure}
\section{RELATED WORK}\label{A}

\textbf{Community Search.} The task of CS aims at the best sub-community in which a given vertex is located [7]. In the present era, contemporary community search models can be classified according to various metrics for evaluating community quality. Examples include \textit{k}-truss [3],  quasi-clique [4], \textit{k}-core [2], and densest subgraph [5]. Furthermore, researchers have explored and introduced novel community models tailored to different large-scale graph data scenarios, encompassing weighted graphs [10], directed graphs [12], and bipartite graphs [13]. Our search model differs from these studies as it involves two query vertices possessing distinct labels and aims to identify a leader pair-based community that integrates two cross-over groups.

\textbf{Attributed to Community Detection and Search.} This line of research  aims to identify communities with a heterogeneous structure and different labels. Graphs for such tasks possess multiple vertex attributes, which differs from our work; the labels in our work cannot directly reflect the interaction relationships between vertices in different communities. [14] proposed attribute-based community search models, which perform well in conducting query tasks on large graphs. However, they identify communities composed of diverse attribute individuals rather than representing cross-community relationships between different communities.

\textbf{Butterfly-Core Community(BCC) Search.} Zheng et al. [1] presented the \textit{Butterfly-Core Community} (BCC) search approach to find cross-group communities encompassing multiple clusters characterized by elevated butterfly degrees and compact diameters. The procedure involves offline index creation using vertex core and butterfly degrees, the local expansion for a initial subgraph $G_{0}$, BCC extraction, and graph diameter reduction. The construction of the butterfly structure and the offline index accelerates the speed of finding cross-community communities in their method, leading to improved performance. However, finding leader vertices and updating vertex distances during graph reduction is time-intensive. Although finding a BCC problem is NP-hard[1], it can take approximately.

\textbf{Random Walk with Restart} \textbf{(RWR).} The \textit{Random Walk with Restart} (RWR) algorithm is a graph-based technique for information dissemination and vertex ranking. It mimics a random walk on a graph wherein, during each step, the walker can potentially reset its path from a predefined set of seed vertices. RWR assigns scores to vertices based on their likelihood of being reached by the random walk. Current RWR-based community search models, such as those in [15], utilize the fine-tuned properties of the RWR algorithm to retain original graph information and enhance the quality of homogeneous community search. However, this contrasts with our goal of seeking communities spanning different clusters.

\section{Preliminaries}\label{A}
In this section, we outline the background, problem definition, and algorithm frame.

\subsection{Labels Graph}\label{AA}
Considering a graph \textit{G = (V, E, \textbf{M})}, where \textit{V} denotes the vertex set, $E \subseteq ${$V\times$V} represents the collection of undirected
and unweighted edges, and \textit{\textbf{M}}: \textit{V}\textrightarrow \textit{\textbf{A}} signifies a vertex label function that maps vertices from \textit{V} to labels in \textit{\textbf{A}}. In a intuitive sense, for every vertex $v\in$\textit{V}, \textit{v} corresponds to an label $\textbf{\textit{M}}(v)\in $\textit{\textbf{A}}. For example, shown in Figure 1, \textit{G} has two labels \{SN, NLP\}. The label of vertex $q_{l}$ is SN,  the edge ($q_{l}$, $a_{1}$) is a homogeneous edge for the label \{SN, SN\}, and the edge ($q_{l}$, $q_{r}$) is a heterogeneous edge for the label \{SN, NLP\}. In this paper, there are two vertex labels.

\subsection{K-Core and Butterfly-Core}\label{AA}
DEFINITION 1 (\textit{k}-CORE). \textit{For a given graph G = (V, E, \textbf{M}), a subgraph S is defined as a k-core if all its vertices have at least k neighbors. Specifically, $S\subseteq$G and ${V_{s}}\subseteq$V, where every vertex in $V_{s}$ possesses at least k neighbors. The core value of a vertex is represented as $\delta$(\textit{v}).}

\

DEFINITION 2 ({\large B}UTTERFLY {\large D}EGREE). \textit{In the context of a bipartite graph B = ($V_{L}$, $V_{R}$, E), with {$E\subseteq${$V_{L}$ × $V_{R}$}}, a butterfly X is defined as a $2\times$2 biclique of G that is formed by four vertices: $v_{l_1}, v_{l_2}\in$$V_{L}$, and $v_{r_1}, v_{r_2}\in$$V_{R}$. This configuration encompasses all four edges ($v_{l_1}$, $v_{r_1}$), ($v_{l_1}$, $v_{r_2}$), ($v_{l_2}$, $v_{r_1}$), and ($v_{l_2}$,$v_{r_2}$), all of which are present in X, constituting a $2\times$2 biclique of G. The butterfly degree of a vertex v is represented as $\chi$(v), signifying the number of butterfly subgraphs in \textit{B} that encompass vertex \textit{v}.}

\subsection{Butterfly-Core Community}\label{AA}

A graph \textit{G} possesses two labels. It contains a butterfly subgraph \textit{B} that is made up of cross-community edges and vertices. a \textit{($k_{1}$,$k_{2}$,\textit{b})}-Butterfly-Core Community (BCC) $G_{H}\subseteq$ \textit{G} adheres to the subsequent criteria:

\textit{1. Different labels: exist two labels {$A_{l}$}$\in$\textbf{\textit{A}}, {$A_{r}$}$\in$\textbf{\textit{A}}, $V_{L}$=\{ $v\in$B: \textbf{\textit{M}}(v)=$A_{l}$\}, $V_{R}$=\{v$\in$B: \textbf{M}(v)=$A{r}$\}. Notice, $V_{L}$ $\cap$$V_{R}$=$\emptyset$, $V_{L}$$\cup$$V_{R}$=$V_{B}$;}
\vspace{0.7ex}

\textit{2. Left core: the subgraph \textit{L} induced by $V_{L}$ is $k_{1}$-core;}

\vspace{0.7ex}

\textit{3. Right core: the subgraph \textit{R} induced by $V_{R}$ is $k_{2}$-core;}

\vspace{0.7ex}

\textit{4. Cross-group interactions: $\exists$$v_{l}$$\in$$V_{L}$, $\exists$$v_{r}$$\in$$V_{r}$ satisfies that butterfly degree $\chi$($v_{l}$)$\geq$b and butterfly degree $\chi$($v_{r}$)$\geq$b.}

\vspace{0.7ex}

\textit{5. Leader Vertex: $\exists$v$\in$$V_{B}$, $\exists$v$\in$$A_{l}$}(\textit{or} $\exists$\textit{v}$\in$$A_{r}$),\textit{ the butterfly degree of vertex satisfy} $\chi$(\textit{v})$\geq$\textit{b}.
\vspace{0.5ex}

\textit{6. Query Distance: The maximum length of the shortest path from v$\in$$V_{G}$ to a query vertex q$\in$Q, i.e., $dist_{G}$($V_{G}$, Q)=$max_{v\in_G, q\in_Q}$}\textit{dist}(\textit{v}, \textit{q}).
\begin{algorithm}
	\caption{Basic Overall Algorithm}
	\begin{algorithmic}[1] 
		\setlength{\baselineskip}{1.5em} 
		\REQUIRE A graph \textit{G} = \textit{L} $\cup$ \textit{R} $\cup$ \textit{B}, \textit{Q} = \{\textit{$q_{l}$, $q_{r}$}\}, three
		
		\hspace{2.3em} integers \{$k_{1}$, $k_{2}$, \textit{b}\}.
		\ENSURE A connected \textit{($k_{1}$,$k_{2}$,\textit{b})}-BCC \textit{$G_H$} containing \textit{Q} 
		
		\hspace{1.8em} with a small diameter.
		
		\STATE \textit{m} $\leftarrow$ 0;
		
		\STATE Compute a shortest path \textit{P} connecting \textit{Q};
		
		\STATE Iteratively expand \textit{P} into graph $G_{m}$ = \{\textit{v} $\in$ \textit{L} $|$ $\delta(\textit{v}) \geq k_l$\} $\cup$ \{\textit{v} $\in$ \textit{R} $|$ $\delta(\textit{v}) \geq k_r$\} by adding adjacent vertices, until $\arrowvert$$G_{m}$$\arrowvert$ $>$ $\eta$;
		
		\STATE Find a maximal connected \textit{($k_{1}$,$k_{2}$,b)}-BCC that includes \textit{Q} within \textit{$G_{H}$}, with the highest coreness on each side.
		
		\STATE \textbf{while} $connected_{G_m}$(\textit{Q}) = \textbf{true} \textbf{do}
		
		\STATE \hspace{2.5em} Compute \textit{dist(v, q)}, \textit{v} $\in$ $V_{G_H}$, \textit{q} $\in$ \textit{Q};
		
		\STATE \hspace{2.5em} vertex $\leftarrow argmax_{v\in{V_{G_m}}}$ \textit{dist(v, q)};
		
		\STATE \hspace{2.5em} Remove the vertex and its edges from \textit{$G_{m}$};
		
		\STATE \hspace{2.5em} Maintain $G_{m}$ as a \textit{($k_{1}$,$k_{2}$,b)}-BCC;
		
		\STATE \hspace{2.5em} $\textit{G}_{m+1}$ $\leftarrow$ $G_{m}$;  \hspace{0.5em}\textit{m} $\leftarrow$ \textit{m} + 1;
		
		\RETURN $G_{H}$ $\leftarrow$ $\arg$ $min_{G^{\prime}\in\{G_{0},...,G_{m-1}\}}$ $dist_{G^{\prime}}(G^{\prime}, Q)$
	\end{algorithmic}
\end{algorithm}
\subsection{Problem Formulation}\label{AA}
We state the BCC-Problem studied in this paper.

\

PROBLEM 1 (BCC-Problem): For a \textit{G} = (\textit{V, E}, \textbf{\textit{M}}), given two query vertices \textit{Q} = \{$q_{l}$,$q_{r}$\} $\in$ \textit{V}, and three integers \{$k_{1}$,$k_{2}$,\textit{b}\}, the BCC-Problem aims to identify a BCC \textit{$G_{H}$} $\subseteq$ \textit{G} that fulfills the subsequent criteria:

1. \textit{Participation} \& \textit{Cohesiveness}: \textit{ The subgraph \textit{$G_{H}$} is a connected ($k_{1}$,$k_{2}$,\textit{b})-BCC with containing \textit{Q} = \{$q_{l}$, $q_{r}$\}}

\vspace{0.4ex}

2. \textit{Connectivity}: \textit{The subgraph \textit{$G_{H}$} is connected}

\vspace{0.4ex}

3. \textit{Most minor diameter}: \textit{The diameter of $G_{H}$ is denoted as diam($G_{H}$). } \textit{$G_{H}$ exhibits the smallest diameter, meaning that} there is no subset {$G_H^{\prime}$}$\subseteq$\textit{$G_{H}$} with \textit{diam}($G_H^{\prime}$)$<$\textit{diam}(\textit{$G_{H}$}), \textit{while} $G_H^{\prime}$ \textit{fulfills the criteria above}.

\vspace{0.6ex}

The BCC-Problem aims to identify a dense Butterfly-Core Community (BCC) with a minor diameter, ensuring minimal communication cost among group members. Furthermore, we eliminate vertices unrelated to the query during the search process.

\

{\large E}XAMPLE 1. \textit{Let the inputs as Q = \{$a_{1}$, $q_{r}$\}, $k_{1}$}=3, \textit{$k_{2}$}=5, \textit{b}=1. \textit{Figure 2 illustrates a} (3, 5, 1)-\textit{BCC}, \textit{where the} 3\textit{-core subgraph L and the} 5\textit{-core subgraph R are evident}. \textit{Subgraph} \textit{B}\textit{ forms a butterfly configuration across} \textit{L} \textit{and} \textit{R}, \textit{and} $\chi$($a_{1}$) = $\chi$($q_{r}$) = 1. \textit{Notice,} \textit{that the vertex} $a_{1}$ \textit{and} $q_{r}$ \textit{is leader vertex}.

\subsection{Basic Overall Solution}\label{AA}
In our previous introductions, we presented a basic overall solution that involves several essential steps: Offline index construction (Each vertex core value in \textit{L} or \textit{R}); Local expansion (lines 2-3); Extract BCC (line 4); Graph diameter reduction (lines 5-10). Finally, return BCC graph $G_{H}$ with minimal diameter.

Algorithm 1 outlines our baseline algorithm framework. BFS expands from vertices on the shortest path between two query vertices during local expansion. To be precise, the procedure initiates with vertices in \textit{P}. These vertices are segregated into $V_{L}$ and $V_{R}$ based on their labels. The minimum coreness of vertices on each side is determined as $k_{l}$ and $k_{r}$. The expansion of \textit{P} is carried out at distinct core values due to the varying densities of the two groups. The vertex limit threshold $\eta$ as an empirical parameter. While reducing graph diameter, maintain BCC structure: Batch remove vertices (\textit{L} or \textit{R}) farthest from the \textit{Q}, update subgraph \textit{B} and leader vertex butterfly degree [1] or refind a new leader vertex for \textit{L} (or \textit{R}); Compute query distance for all vertices in the current graph $G_{m}$. We select the highest butterfly degree as the subgraph \textit{L} (or \textit{R}) leader vertex.

\subsection{Analysis Basic Overall Algorithm}\label{AA}
While reducing the graph diameter triggers updates to the leader vertex, finding the leader vertex with the highest butterfly degree in \textit{B} requires calculating the butterfly degree for all vertices in \textit{B} (In the worst case), which is time-consuming. We consider the significance of vertices within the current community and modify the strategy for updating leader vertices to achieve acceleration. Furthermore, calculating query distances for all vertices after each vertex deletion is time-consuming due to repeated shortest-path calculations. Drawing inspiration, we update query distances for all vertices using the BFS method only once.

\section{Design of Fast-BCC Model}\label{A}
\subsection{Comprehensive Vertex Scoring Method}\label{AA}
We introduce a novel comprehensive vertex scoring method based on vertex importance within homogeneous communities and their butterfly degrees. The method employs the \textit{Random Walk with Restart} (RWR) algorithm to calculate all vertices scores from a seed vertex, with higher scores indicating greater significance. To incorporate butterfly degrees, we provide the formula for the RWR scores:
\[ \hspace{3em}Vec(v_t) = ( scores_{v_{t1}} \ldots scores_{v_{tt}} \ldots scores_{v_{ti}} ) \hspace{1em}(1)
\]
\textit{Vec}($v_{t}$) is the vertex score vector obtained by running the RWR algorithm with $v_{t}$ as the seed vertex. It encompasses the intimacy scores of each vertex \textit{v} in \textit{G} to the vertex $v_{t}$, where $v_{t}$ also holds an intimacy score with itself.

\begin{figure}[htb]
	\centering
	\includegraphics[width=0.47\textwidth, height=0.15\textheight]{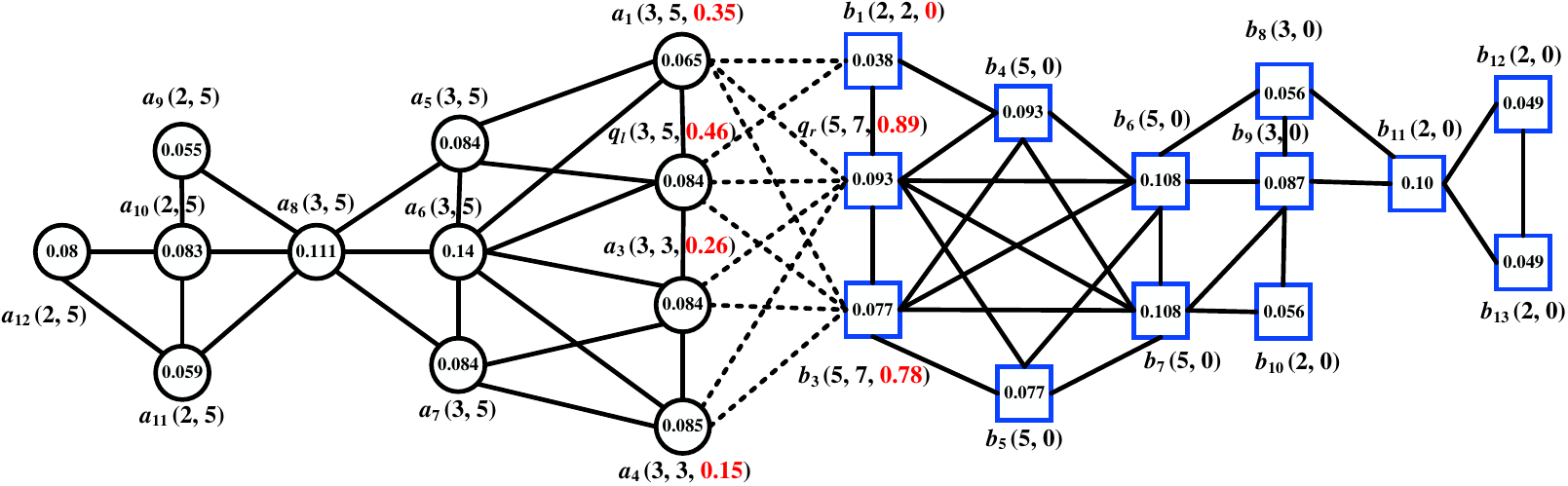}
	\caption{An example of labeled \textit{G}, each vertex note as $v_{i}$( $\delta$($v_{i}$), $\chi$($v_{i}$), \textit{VSC}($v_{i}$) ). The numbers within the graph represent RSN($v_i$) scores.}
	\label{fig:figure3}
\end{figure}

\[ \hspace{8em}RS(v_t) = \frac{\sum_{i=1}^{\arrowvert V \arrowvert} scores_{v_{ti}}}{\arrowvert V \arrowvert} \hspace{5.5em}(2)
\]

For $v_{t}$$\in$\textit{V}, we aggregate the $scores_{v_{ti}}$ obtained by each vertex across all runs \textit{Vec}(\textit{v}) and calculate the average to determine the relative average score of each vertex to \textit{G}, where $\arrowvert V$$\arrowvert$ is denoted by the vertex count.

\[ \hspace{6em}RSN(v_t) = \frac{ RS(v_t) -  RS(v)_{min} }{RS(v)_{max} - RS(v)_{min}}
\hspace{2.5em}(3)\]
$RS_{max}(v)$ denotes the maximal \textit{RS}(\textit{v}) and $RS_{min}(v)$ denotes the minimum \textit{RS}(\textit{v}). \textit{RSN} ensures that each vertex's scores fall within the range of [0,1].

\

DEFINITION 3 ({\large V}ERTEX {\large B}UTTERFLY {\large S}CORES) \textit{Given a bipartite graph B, a set vertices $V_B$. Vertex butterfly scores are defined as}:

\hspace{6em}$\displaystyle BSN(v_t)=\frac{\chi(v_t) - \chi(v)_{\min}}{\chi(v)_{\max} - \chi(v)_{\min}}$ \hspace{3em}(4)

\

For v$\in$$V_B$, we compute vertex butterfly degree$\chi$(\textit{v}). We normalize all vertex's butterfly degree $\chi$(\textit{v}), which can ensure that each vertex's butterfly scores fall within the [0, 1] range.

\

DEFINITION 4 ({\large V}ERTEX {\large C}PREHENSINVE {\large S}CORES) \textit{For a graph G and vertex set V, the vertex comprehensive scores are defined as: }

\

\hspace{4.5em} $VSC(v_t) = \gamma_{1}RSN(v_t) + \gamma_{2}BSN(v_t)$ \hspace{1.5em}(5)
\vspace{0.5ex}

Vertices with comprehensive scores encompass their intimacy with other vertices in the homogenous graph and their interactive relationships with vertices in the heterogeneous graph. The parameters $\gamma_{1}$ and $\gamma_{2}$ are empirical. Note that we do not consider vertices with a degree of 0 in the computation of the comprehensive scores.

\subsection{Fast Identify Leader Vertex Method (FILVM)}In the following section, we present two observations that assist us in identifying a more effective pair of leader vertices from [1] and Figure 3.

\

{\large O}BSERVATION 1. In the community, the leader vertex combines high RWR scores for effective intra-community connections and a substantial butterfly degree for inter-community links. Consequently, the leader vertex's resilience in the graph operations increases.

\

{\large O}BSERVATION 2. Leader vertices show a low query distance, signifying their closeness to query vertices and making them less prone to removal in graph operations.

\

\textbf{Algorithm FILVM.} We proposed a method for fast-finding and updating leader vertices. Our search scope includes vertices within the current subgraph that possess comprehensive scores. When finding a vertex meets the butterfly degree greater than or equal to b, it is immediately returned as a leader vertex. Specifically, if the query vertex meets the butterfly degree criteria, it becomes the leader vertex. Otherwise, we explore vertices with comprehensive scores, starting with the query's 1-hop neighbors. If no suitable leader is found, we continue searching among other vertices. Evaluating vertices in descending order of comprehensive scores, we identify qualifying leaders. If still unsuccessful, the query vertex is used as the leader.

\

\begin{figure}[htb]
	\centering
	\includegraphics[width=0.47\textwidth, height=0.15\textheight]{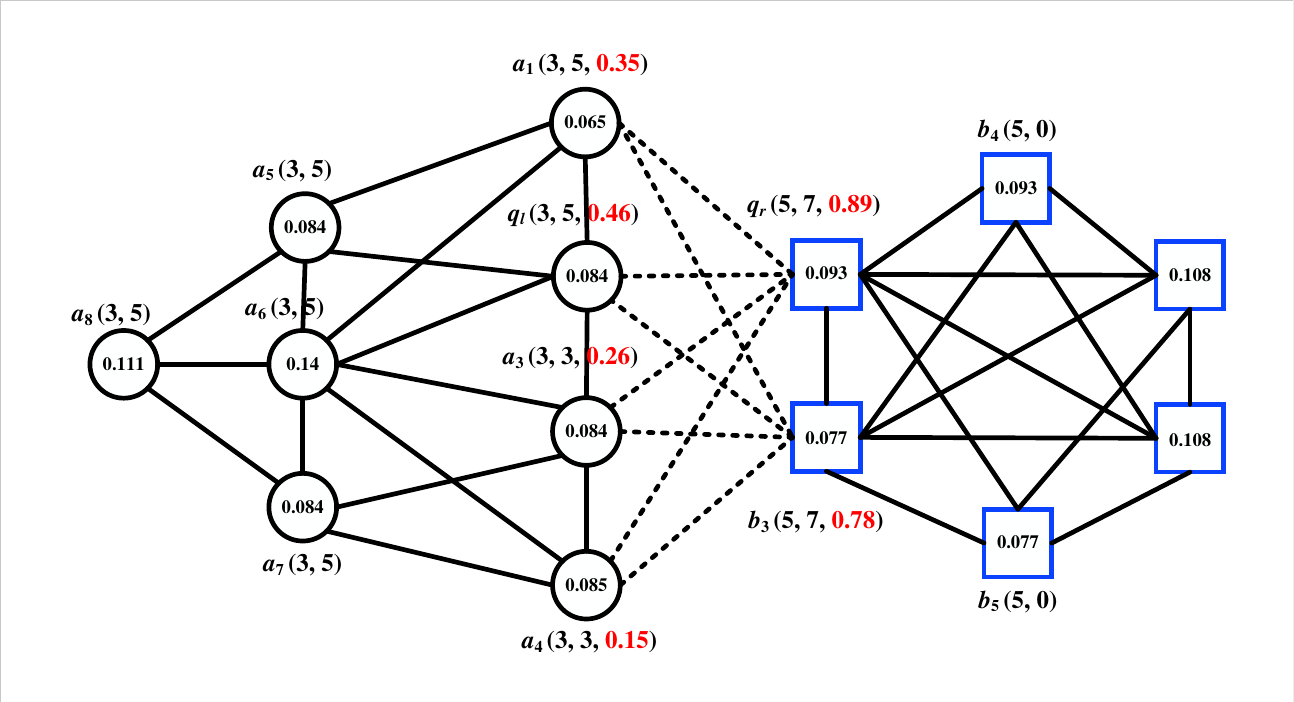}
	\caption{Given \textit{Q} = \{$a_{7}$, $b_{4}$\}, a connected BCC \textit{$G_0$}, each vertex note as $v_{i}$($\delta$($v_{i}$), $\chi$($v_{i}$), VSC($v_{i}$)). The numbers within the graph represent RSN($v_i$) scores.}
	\label{fig:labelname}
\end{figure}

{\large E}XAMPLE 2. \textit{Given a Q = \{$a_{7}$, $b_{4}$\}. Our approach computes the butterfly degree once. Specifically, above previous introductions $G_0$ shown in Figure }4, \textit{calculating a butterfly degree of} 3 \textit{for vertex $a_{3}$ and designating $a_{1}$ as the leader vertex. In contrast, the BCC model calculates the butterfly degree for vertices \{$a_{1}$, $q_{l}$, $a_{3}$, $a_{4}$\} four times to choose the vertex as leader with a relatively high degree and are }3-\textit{hop neighbors of the query. }

\
As the BCC model employs a 3-\textit{hop} greedy algorithm to search for vertices with maximized butterfly degree, when the extremum values of butterfly degree within the community are relatively high, such repeated 3-\textit{hop} iterations can lead to an increased runtime. Moreover, we are not limited to a 3-hop range when updating the leader vertex. As long as there are leader vertices that meet the criteria within the current community, we can capture a BCC structure.

\begin{algorithm}[h]
	\caption{\textit{BQDC}($G_{i}$, \textit{q}, $V_{D_{i}}$)}
	\begin{algorithmic}[1] 
		\setlength{\baselineskip}{1.5em} 
		\REQUIRE Graph \textit{$G_{i}$}, query vertex \textit{q}, removal vertices $V_{D}$
		\ENSURE The updated distance dist(\textit{v}, \textit{q}) for all vertices \textit{v}.
		
		\STATE Delete vertices $V_{D_{i}}$ and their incident edges from $G_{i}$
		
		\STATE \textit{path} $\leftarrow$ 1
		
		\STATE Initialize a visited vertices set \textit{C} as empty;
		\STATE Initialize a queue \textit{A} and a hashmap \textit{K} as empty;
		\STATE Add \textit{q} into \textit{C}, add \textit{q} into \textit{A};
		
		\STATE \textbf{while} {\textit{A} is not empty} \textbf{do}
		
		\STATE \hspace{0.65em} \textbf{for all} {range(0, $|A|$)} \textbf{do}
		
		\STATE \hspace{1.65em} Pop vertex \textit{v};
		
		\STATE \hspace{1.65em} new visited vertices set $N_{new}$ $\leftarrow$ (\textit{N}(\textit{v}) $\textbf{\textendash}$ \textit{C});
		
		\STATE \hspace{1.65em} \textbf{for all} {vertex \textit{u} in $N_{new}$  $\cap$ $V_{G_i}$} \textbf{do}
		
		\STATE \hspace{2.7em} \textit{K}(\textit{u}) $\leftarrow$ \textit{path};
		
		\STATE \hspace{1.65em} Add \textit{u} in $N_{new}$ into \textit{C} and \textit{A};
		
		\STATE \hspace{0.65em} \textit{path} $\leftarrow$ \textit{path} + 1;
		
		\RETURN \textit{K};
	\end{algorithmic}
\end{algorithm}

\begin{algorithm}[ht]
	\caption{Fast-BCC Model}
	\begin{algorithmic}[1] 
		\setlength{\baselineskip}{1.5em} 
		\REQUIRE A graph \textit{G} = \textit{L} $\cup$ \textit{R} $\cup$ \textit{B}, \textit{Q} = \{\textit{$q_{l}$, $q_{r}$}\}, three
		
		\hspace{2.3em} integers \{$k_{1}$, $k_{2}$, \textit{b}\}.
		\ENSURE A connected \textit{($k_{1}$,$k_{2}$,\textit{b})}-BCC \textit{$G_H$} containing \textit{Q} 
		
		\hspace{1.8em} with a small diameter.
		
		\STATE Compute a shortest path \textit{P} connecting \textit{Q};
		
		\STATE $k_{l}$ $\leftarrow$ $min_{v \in V_{L}}\delta(\textit{v})$;
		$k_{r}$ $\leftarrow$ $min_{v \in V_{R}}\delta(\textit{v})$;
		
		\STATE Iteratively expand \textit{P} into graph \textit{$G_H$} = \{\textit{v} $\in$ \textit{L} $|$ $\delta(\textit{v}) \geq k_l$\} $\cup$ \{\textit{v} $\in$ \textit{R} $|$ $\delta(\textit{v}) \geq k_r$\} by adding adjacent vertices, until $\arrowvert$$G_{H}$$\arrowvert$ $>$ $\eta$;

		\STATE Find a connected \textit{($k_{1}$,$k_{2}$,b)}-BCC containing \textit{Q} of \textit{$G_{H}$} with the largest coreness on each side;
		
		\STATE Remove disqualified subgraphs on $G_{H}$ with Algorithm 
		\hspace{1.8em}FILVM and Algorithm 2 to return the final BCC;
	\end{algorithmic}
\end{algorithm}

\subsection{BFS-Query Distance Computation (BQDC)}\label{AA}
We introduce a novel method for updating vertex distances. We skip the step of checking distance relationships between vertices. Instead, we directly initiate a BFS from the query vertex to expand throughout the current subgraph, thereby updating the distances between vertices.

\textbf{Algorithm.} First, remove the vertices farthest from \textit{Q} in the graph $G_{i}$ (line 1). Next, initialize parameters (lines 2-5). Lines 6-13 detail the BFS algorithm. After that the first visit to vertex \textit{v} in graph $G_{i}$, we update the distance between \textit{v} and \textit{q} as a path and store it in the hashmap \textit{K}. Once all vertices previously enqueued are dequeued, increment the path by 1 (line 13), and then start a new iteration until the queue is empty. Finally, return the updated query distance hashmap \textit{K} (line 14).

Although BFS may involve repeated vertex visits, in a \textit{k}-core community, vertex distances are not notably distant. This acceleration method remains viable and has been verified in our experiments.

\subsection{Fast-Butterfly-Core Community (Fast-BCC) Model}
We present the framework of our Fast-BCC method, equipped with FILVM, and Algorithm 2 involves several essential steps: Offline index construction (Vertex Comprehensive scoring method); Local expansion (lines 1-3); Extract BCC (line 4); Graph diameter reduction (line 5), in this process, we employ FILVM and Algorithm 2 for acceleration while maintaining consistency with the Basic method in the remaining steps such as updating B, leader, and their butterfly degrees. Finally, return BCC graph $G_{H}$ with minimal diameter.

Although Algorithm 3 does not maintain a 2-approximation guarantee[1] for optimal solutions, it efficiently produces satisfactory outcomes in practice, substantiated by our experimental result.

\begin{table}[h]
	\centering
	\renewcommand{\arraystretch}{1.2} 
	\caption{Network Statistics K=$\textbf{10}^{\textbf{3}}$, M=$\textbf{10}^{\textbf{6}}$}
	\begin{tabular*}{0.45\textwidth}{c|ccccc}
		\hline
		\textbf{Network} & $| V |$ & $| E |$ & \textit{Labels} & $k_{max}$ & $diameter_{max}$ \\
		\hline
		DBLP & 317\textbf{K} & 1\textbf{M} & 2 & 113 & 342 \\
		Amazon & 335\textbf{K} & 926\textbf{K} & 2 & 6 & 549 \\
		Orkut & 3.1\textbf{M} & 117\textbf{M} & 2 & 5 & 33313 \\
		Youtube & 1.1\textbf{M} & 3\textbf{M} & 2 & 51 & 28754 \\
		LiveJournal & 4\textbf{M} & 35\textbf{M} & 2 & 360 & 14815 \\
		\hline
	\end{tabular*}
	\label{tab:network}
\end{table}

\section{Experiments}\label{A}
\subsection{Datasets.}\label{AA}
We use graphs from SNAP (shown in TABLE I) with ground-truth communities, randomly assigning two labels. (\textit{Especially for the LiveJournal and Orkut datasets, we initially sampled communities randomly from the ground-truth data to form connected graphs with 200,000 to 300,000 vertices, constrained by the limitations of the hardware's memory size.}) Split vertices into two labeled communities, connected by 10\% cross edges per community to simulate collaboration. Adding 10\% cross-noise edges introduces noise. Generate query vertex pairs with distinct labels. Since vertex comprehensive scores remain constant in the community searching, we treat RWR algorithm execution and butterfly degree computation as preprocessing steps. Experiments ran on Python 3.6, i5-12400F CPU, and 32GB RAM.

\subsection{Compared Methods.}\label{AA}
As described below, we compare three BCC search techniques against two competing community search methods.

$\bullet$ Basic: Our basic solution is described in Algorithm 1, in which we adopt the maximal diameter in the Database for local expansion paths as used in the original BCC model.

$\bullet$ LP-BCC [1]: Finding a connected Butterfly-Core Community with a small diameter incorporates efficient query distance computation and leader pair identification as presented by LP-BCC.

$\bullet$ Fast-BCC: Our RWR-based Butterfly-Core Community search is described in Algorithm 3, equipped with Algorithm 2, Fast Identify Leader Vertex Method.

Note that our methods employ bulk deletion, which involves removing a batch of vertices with the most tremendous distances from the graph. We use the default parameter setting of BCC. For Fast-BCC, we set the parameters $\gamma_{1}$=0.5, $\gamma_{2}$=0.5. We provided the parameters for the RWR algorithm: restart probability set as 0.15, convergence threshold set as 1e-6, and maximum number of iterations range from 100 to 200.

\textbf{Query vertices and evaluation metrics.} We create query \textit{Q} = \{$q_{l}$, $q_{r}$\} with coreness values $k_{1}$, $k_{2}$ for $q_{l}$ and $q_{r}$, and set \textit{b} to 1. For efficiency, we use seconds to measure method runtimes. Queries taking over 30 minutes are treated as infinite time. F1-score evaluates community quality against ground truth.

\begin{figure}[htb]
	\centering
	\includegraphics[width=0.45\textwidth, height=0.15\textheight]
	{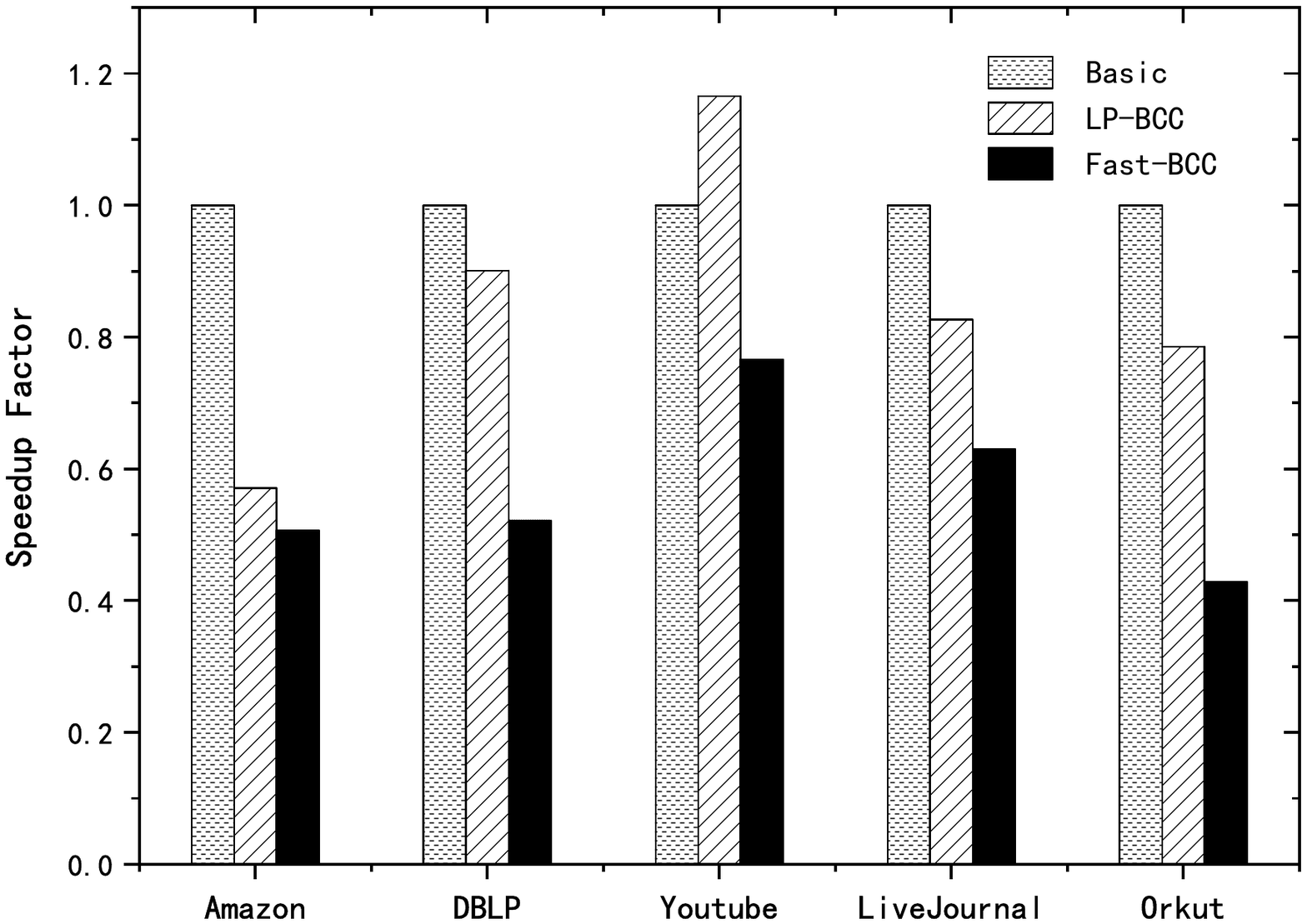}
	\caption{Assessment of efficiency on ground-truth networks}
\end{figure}

\begin{table}[ht]
	\centering
	\caption{Running Time (in Seconds).}
	\begin{tabular}{cccccc}
		\toprule
		\textbf{Network}& Amazon & DBLP & Youtube & LiveJournal & Orkut \\
		\midrule
		Basic & 0.091s & 0.057s & 0.071s & 1.511s & 0.438s \\
		LP-BCC & 0.052s & 0.053s & 0.083s & 1.249s & 0.344s \\
		Fast-BCC & \textbf{0.046s} & \textbf{0.030s} & \textbf{0.054s} & \textbf{0.952s} & \textbf{0.187s} \\
		\bottomrule
	\end{tabular}
	
	\label{tab:mytable}
\end{table}

\begin{figure}[htb]
	\centering
	\includegraphics[width=0.48\textwidth, height=0.15\textheight]
	{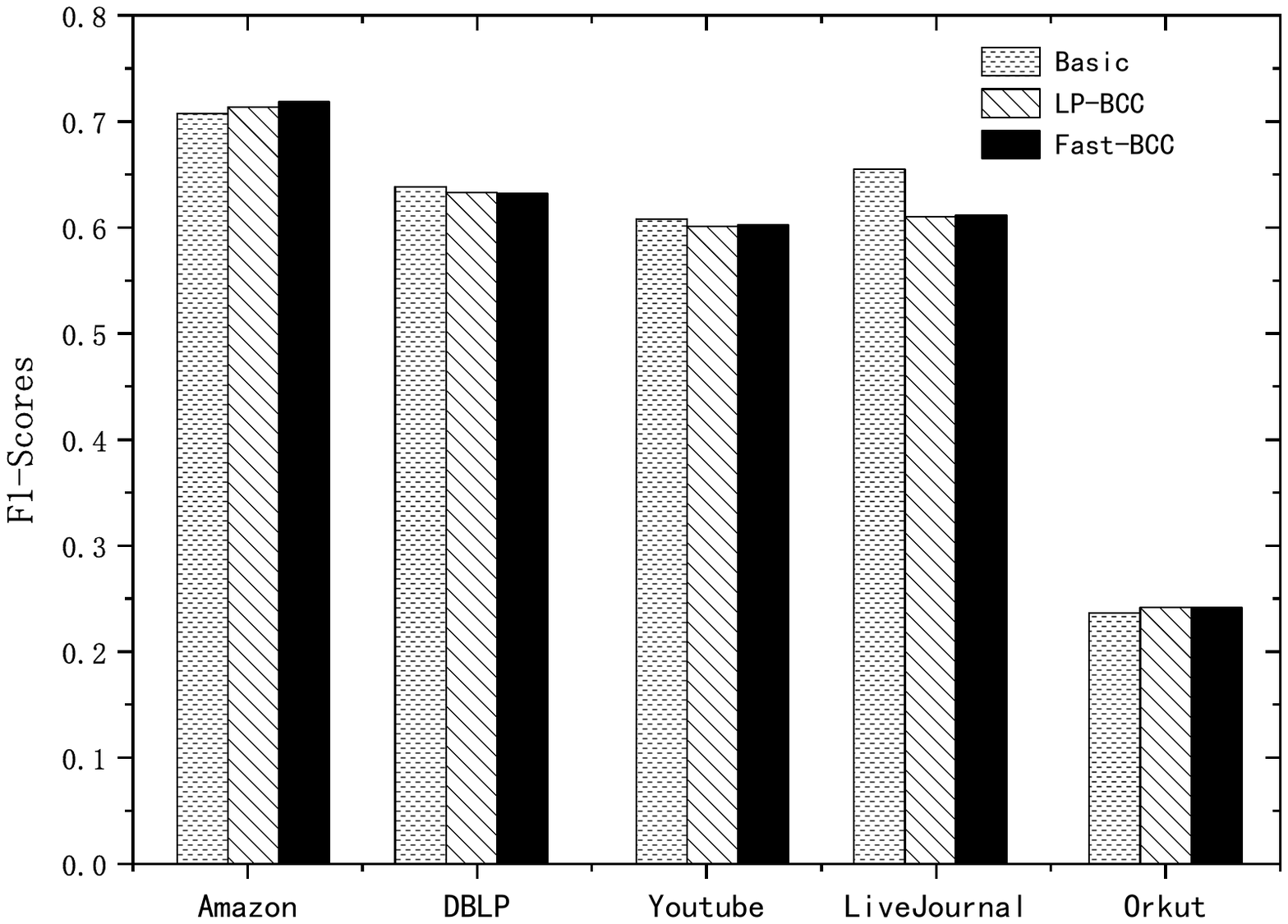}
	\caption{Quality evaluation on ground-truth networks}
\end{figure}

\textbf{Exp-1.} \textbf{Assessment of efficiency for all methods.} We assessed execution speeds across diverse community search models. We generated 1000 query vertex pairs for each method and calculated average execution times. TABLE II shows runtime results. To illustrate method efficiency, we calculated speedup ratios relative to the Basic method runtime for each dataset. Results are shown in Figure 5. Fast-BCC is over 30\% faster than Basic and 10\%-40\% faster than LP-BCC, except in sparse cases like the Amazon dataset. Overall, Fast-BCC outpaces other methods in speed.

\begin{figure}[htbp]
	\centering
	
	\begin{subfigure}[b]{0.15\textwidth}
		\centering
		\includegraphics[width=\textwidth]{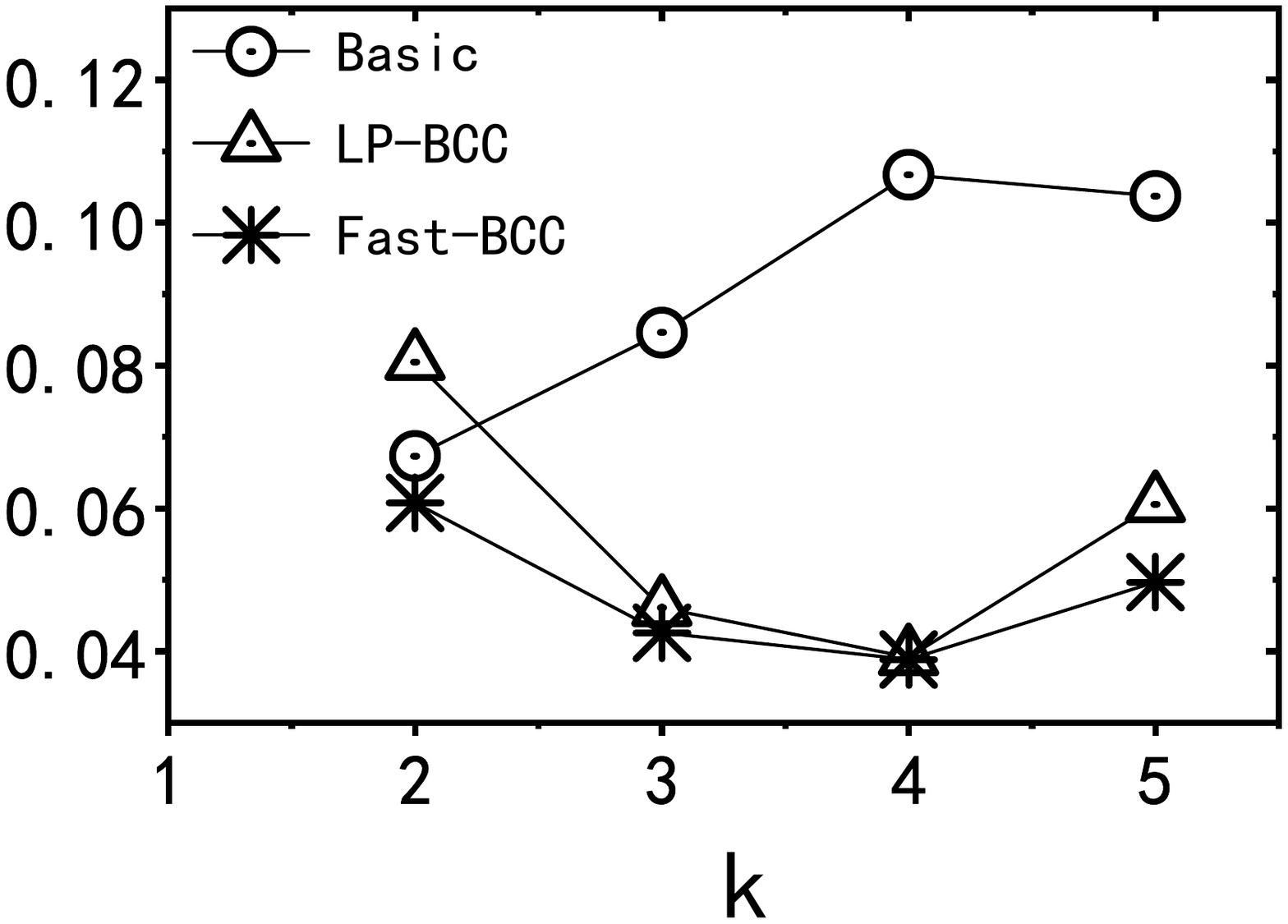}
		\caption{Amazon}
	\end{subfigure}
	\hfill
	\begin{subfigure}[b]{0.15\textwidth}
		\centering
		\includegraphics[width=\textwidth]{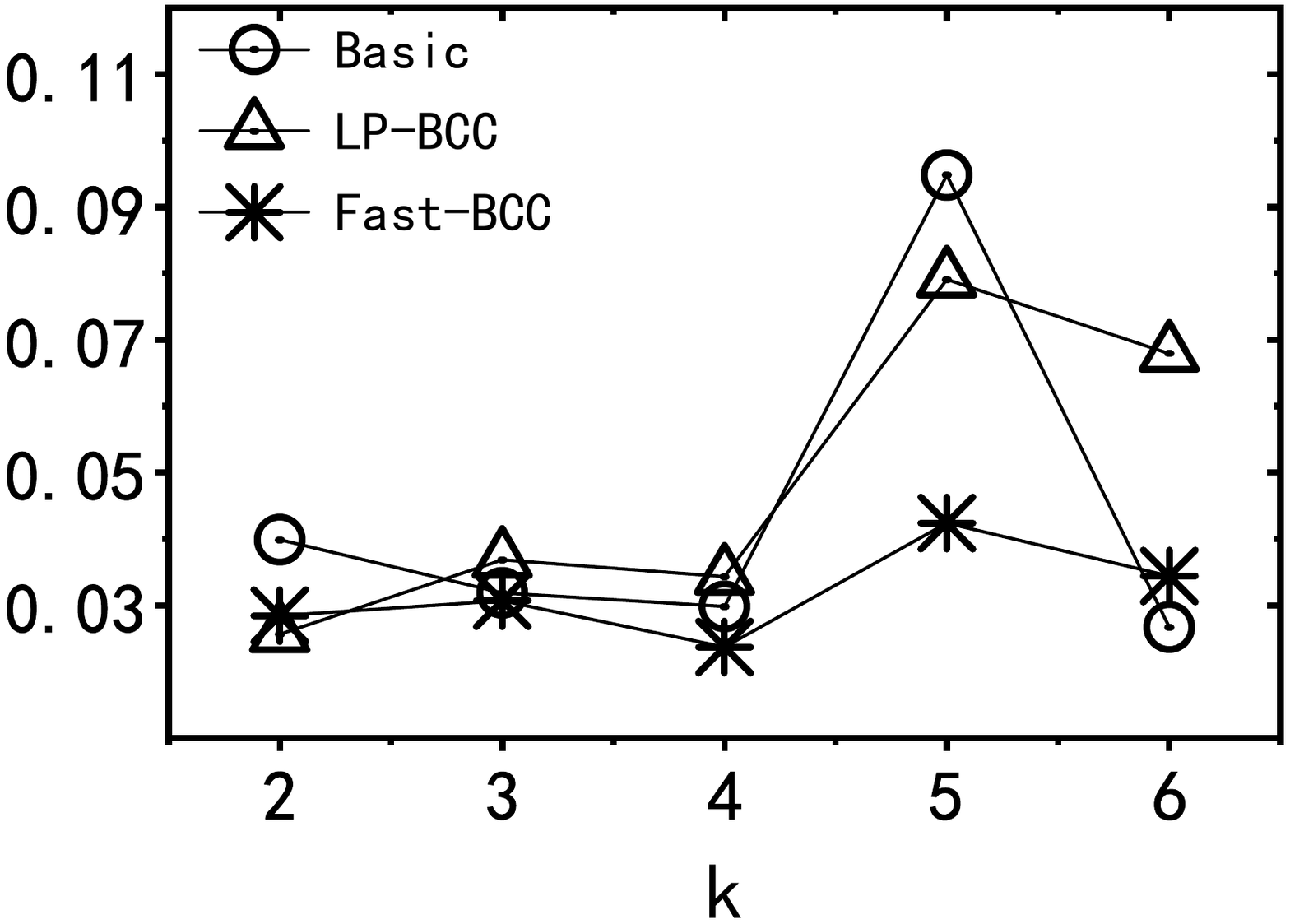}
		\caption{DBLP}
	\end{subfigure}
	\hfill
	\begin{subfigure}[b]{0.15\textwidth}
		\centering
		\includegraphics[width=\textwidth]{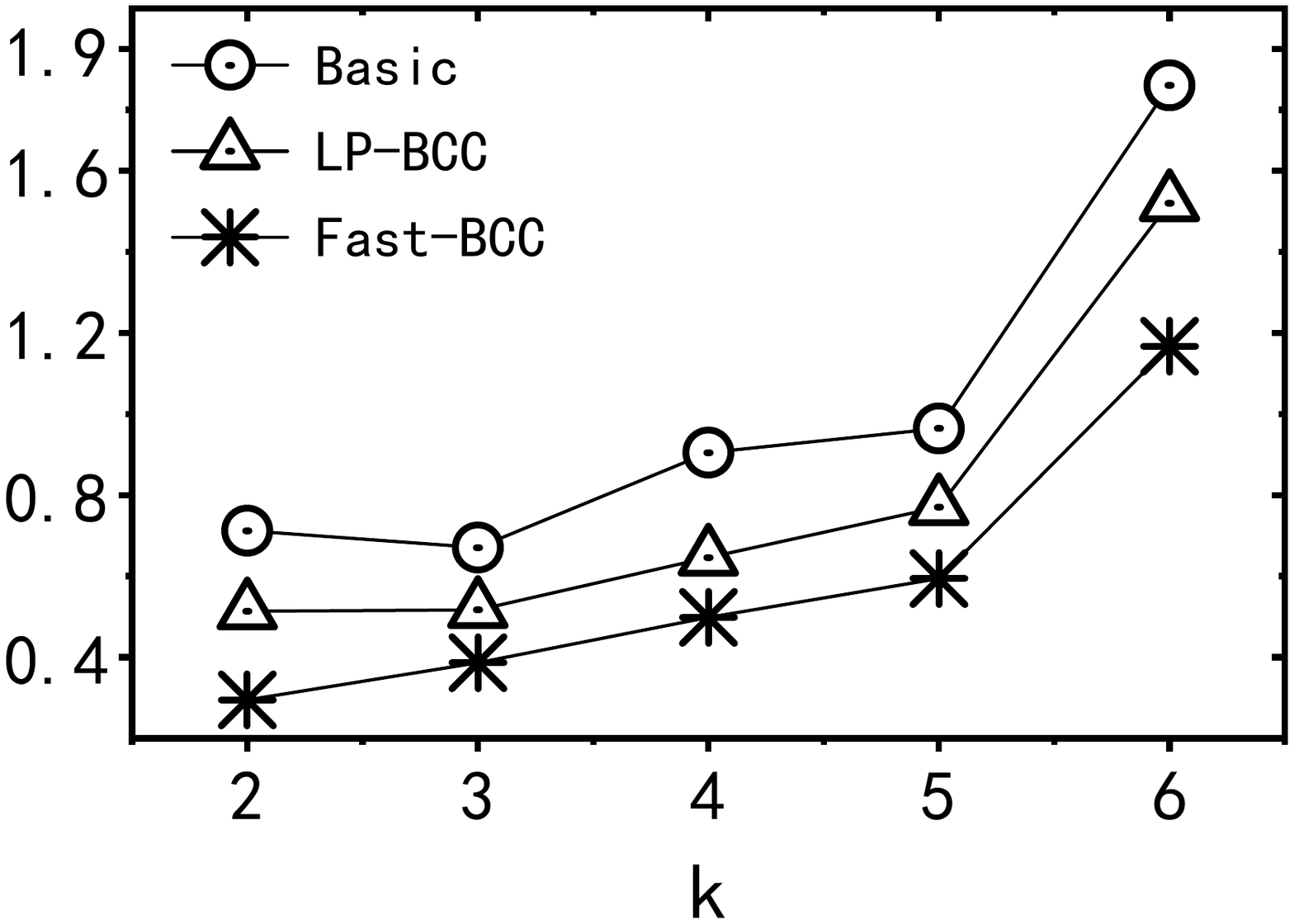}
		\caption{LiveJournal}
	\end{subfigure}
	
	\caption{Compare query time by varying value k.}
	\label{fig:vary_k}
\end{figure}

\begin{figure}[htbp]
	\centering
	
	\begin{subfigure}[b]{0.15\textwidth}
		\centering
		\includegraphics[width=\textwidth]{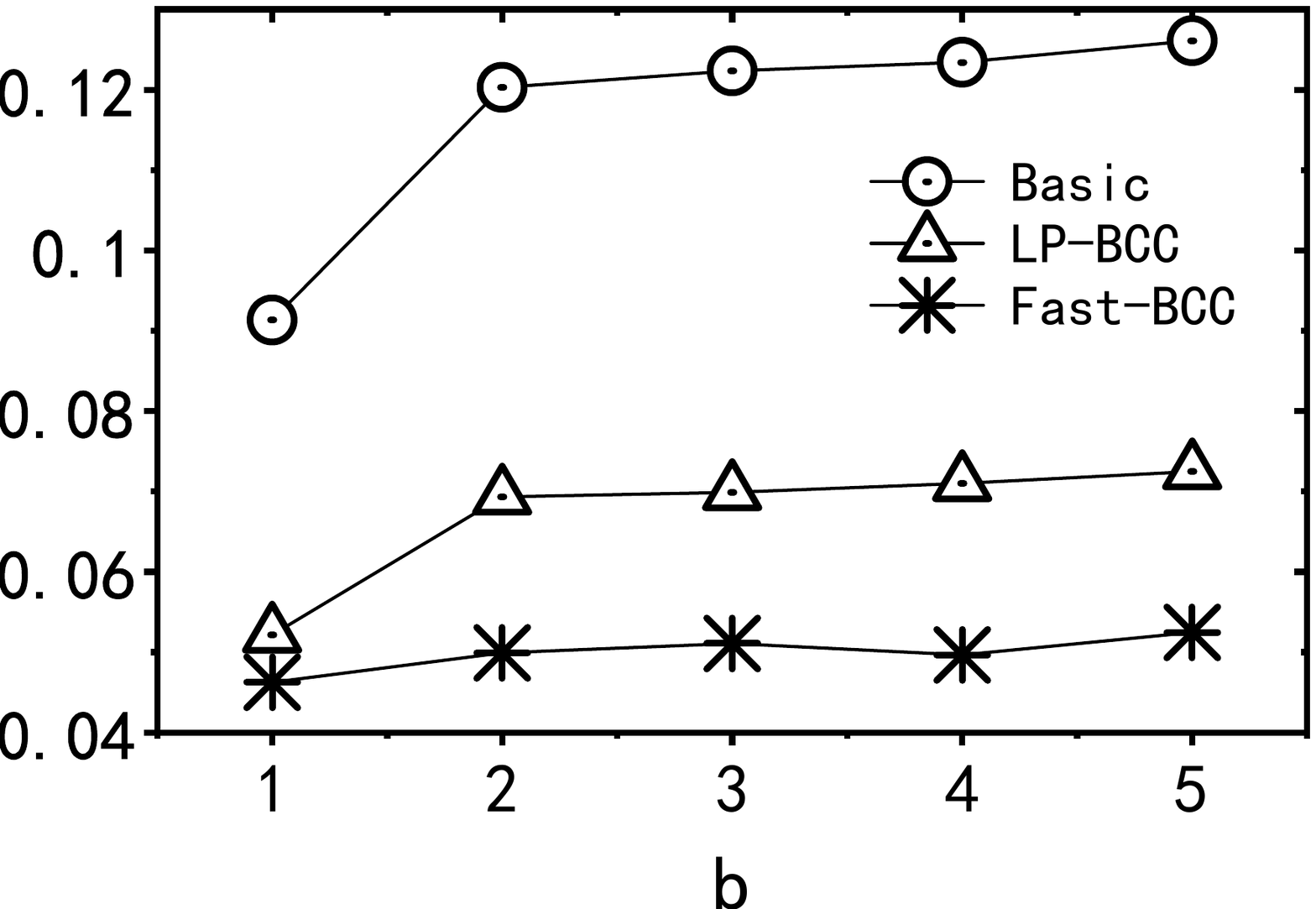}
		\caption{Amazon}
	\end{subfigure}
	\hfill
	\begin{subfigure}[b]{0.15\textwidth}
		\centering
		\includegraphics[width=\textwidth]{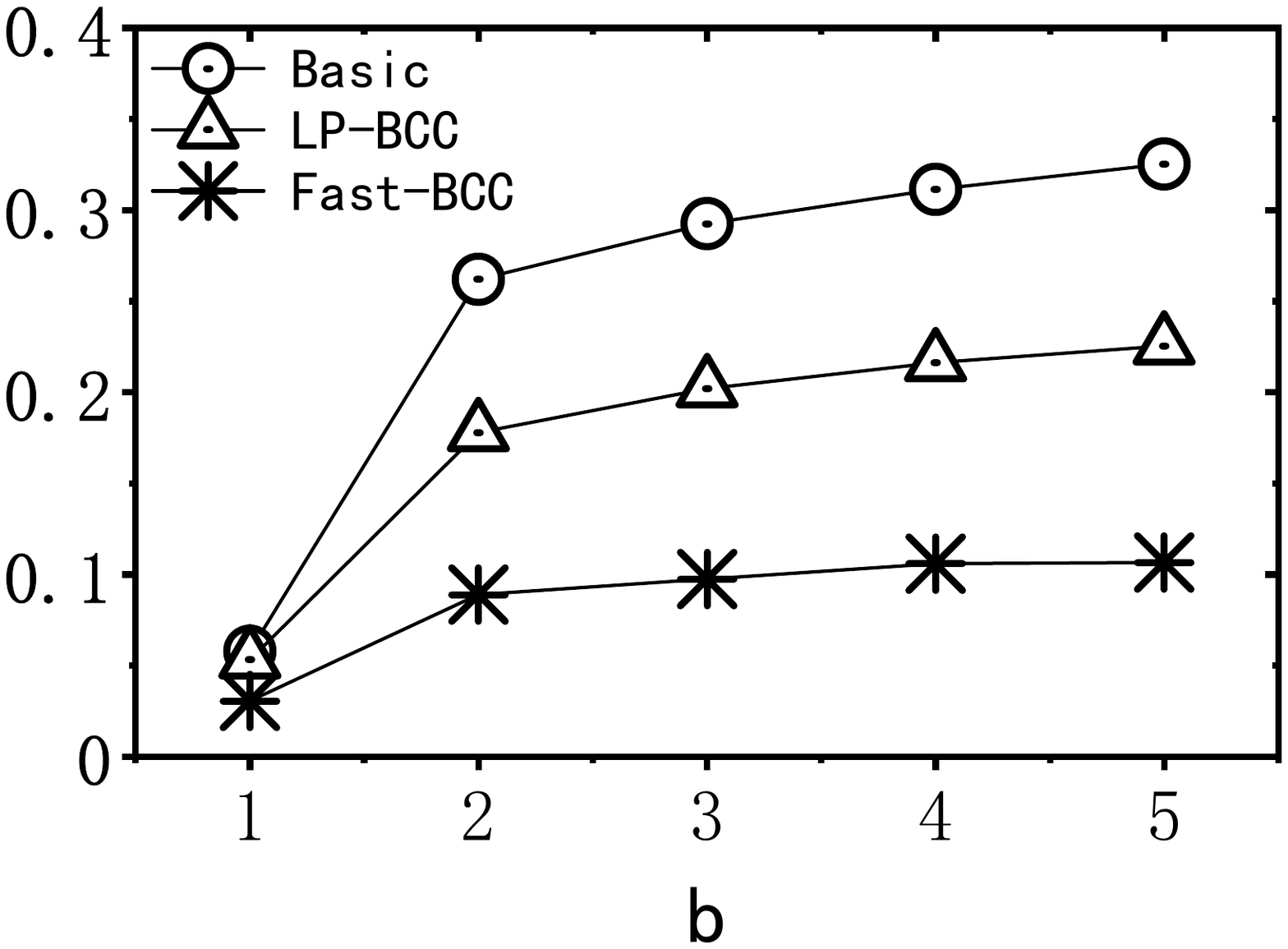}
		\caption{DBLP}
	\end{subfigure}
	\hfill
	\begin{subfigure}[b]{0.15\textwidth}
		\centering
		\includegraphics[width=\textwidth]{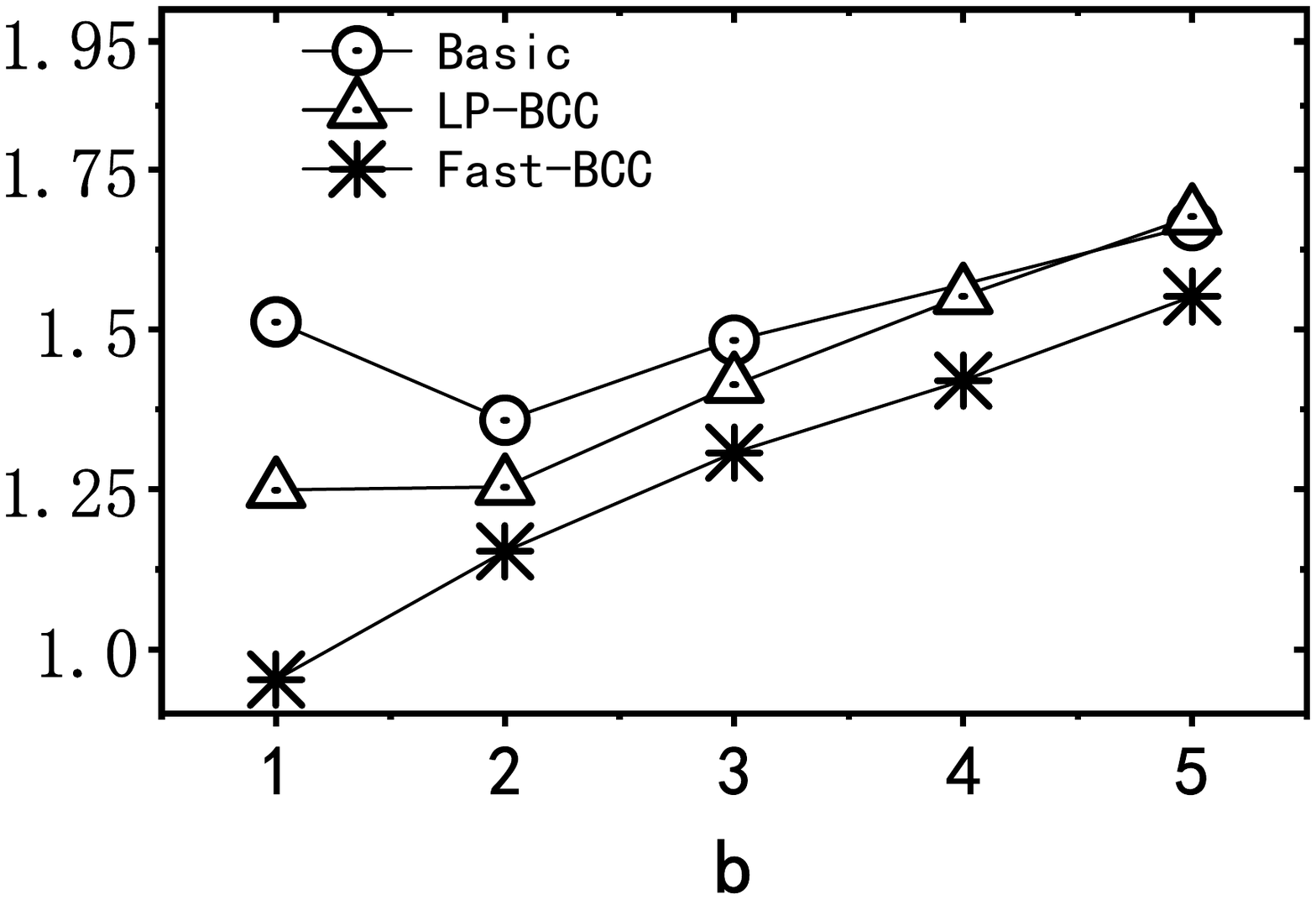}
		\caption{LiveJournal}
	\end{subfigure}
	
	\caption{Compare query time by varying value b.}
	\label{fig:vary_b}
\end{figure}

\textbf{Exp-2.} \textbf{Quality evaluation with ground-truth communities.} We evaluated the efficiency of various community search models on five datasets. We randomly generate 1000 query vertex pairs for each—Figure 6 shows the average F1 scores of different methods across datasets. As shown, Fast-BCC boosts speed and nearly matches the top F1-Score. Orkut dataset posed challenges for all methods.

\textbf{Exp-3.} \textbf{Parameter sensitivity evaluation.} We evaluate the parameter sensitivity of $k_{1}$, $k_{2}$ and \textit{b} on efficiency performance. During the testing of one parameter, the other two parameters remain constant. Moreover, due to their symmetry property, we test one parameter of $k_{1}$ and $k_{2}$. Figure 7 shows runtime changes based on varying the \textit{k} value. Greater \textit{k} values lead to shorter runtimes due to the smaller $G_0$ generated. The Fast-BCC method exhibits optimal efficiency across core values, with efficiency initially dropping and then recovering with higher \textit{k}. Across different \textit{b} values shown in Figure 8, Fast-BCC maintains relatively consistent runtime speeds across the three datasets. The substantial increase in runtime for all methods on LiveJournal stems from its densely populated communities.

\begin{table}[h]
	\centering
	\renewcommand{\arraystretch}{1.3} 
	\caption{Reduction efficiency comparison on LiveJournal regarding query distance calculation and butterfly computation.}
	\begin{tabular}{c|ccc}
		\hline
		Methods & Basic & LP-BCC & Fast-BCC \\
		\hline
		Leader upgrade times & 215 & \textbf{208} & 250 \\
		Leader upgrade cost & 110.45s & 142.99s & \textbf{13.94s} \\
		Query distance computation & 296.94s & 82.11s & \textbf{41.92s} \\
		\hline
	\end{tabular}
\end{table}

\textbf{Exp-4.} \textbf{Efficiency evaluation of the reduction process.} We evaluate the efficiency of our leader pair identification in FILVM and fast query distance computation in Algorithm 2. As shown in Table III, although our Fast-BCC method increases the number of leader updates, it accelerates the updates of leader vertices by 7.9\textbf{x} compared to the Basic method and 10.2\textbf{x} compared to the LP-BCC method. For query distance updates, the Fast-BCC method accelerates by 7\textbf{x} compared to the Basic method and almost 2\textbf{x} compared to the LP-BCC method.

\section{Conclusion}\label{A}
In this paper, we introduce a fast method for locating Butterfly-Core Communities. Leveraging RWR scores and butterfly degrees, we create a novel offline index to evaluate vertex importance within communities. The offline index leads to efficient leader vertex updates and rapid BFS-based vertex distance updates. Moreover, by incorporating RWR scores, we design weighted expansion paths and adapt local expansions to diverse subgraphs. Experiments demonstrate our method's efficacy on extensive real-world community datasets.

\section*{Acknowledgment}
This work is supported by the National Natural Science Foundation of China (Grant No. 61802444), the Changsha Natural Science Foundation (Grant No. kq2202294), the Research Foundation of Education Bureau of Hunan Province of China (Grant No. 20B625, No. 18B196, No. 22B0275); the Research on Local Community Structure Detection Algorithms in Complex Networks (Grant No. 2020YJ009).

\end{document}